\documentclass[prl,twocolumn,superscriptaddress,showpacs]{revtex4}
\usepackage{amsmath}
\usepackage[dvips]{graphicx}
\usepackage[usenames]{color}

\begin{document}

\newcommand{\vc}{\mathbf}
\newcommand{\bK}{{\bf{K}}}
\newcommand{\bS}{{\bf{S}}}
\newcommand{\bk}{{{\bf{k}}}}
\newcommand{\bq}{{\bf{q}}}
\newcommand{\bQ}{{\bf{Q}}}
\newcommand{\br}{{\bf{r}}}
\newcommand{\ha}{{\hat{a}}}
\newcommand{\hb}{{\hat{b}}}
\newcommand{\hc}{{\hat{c}}}
\newcommand{\upa}{{\uparrow}}
\newcommand{\dna}{{\downarrow}}
\newcommand{\nn}{{\nonumber}}
\newcommand{\be}{\begin{equation}}
\newcommand{\ee}{\end{equation}}
\newcommand{\ra}{\rangle}
\newcommand{\la}{\langle}
\newcommand{\bea}{\begin{eqnarray}}
\newcommand{\eea}{\end{eqnarray}}
\newcommand{\pll}{\parallel}

\title{ Temperature dependent Fermi arcs in the normal state of the
underdoped cuprate superconductors}
\author{Arun Paramekanti}
\affiliation{Department of Physics, University of Toronto, Toronto,
Ontario M5S-1A7, Canada}
\author{Erhai Zhao}
\affiliation{Department of Physics, University of Toronto, Toronto,
Ontario M5S-1A7, Canada}
\begin{abstract}
Angle resolved photoemission experiments by Kanigel, et al [Nature
Physics 2, 447 (2006)] have made a remarkable observation that low energy
electronic excitations in the normal state of underdoped cuprate
superconductors exist on open ``Fermi arcs'' instead of a closed
Fermi surface. These arcs shrink upon cooling, with the arc length
appearing to extrapolate to nodal 
points at zero temperature. We show that this striking non-Fermi liquid 
behavior could result from the underdoped normal state above $T_c$
lying in the vicinity of a quantum phase transition between a d-wave 
superconductor and a correlated insulating phase.
\end{abstract}
\maketitle

{\it Introduction. ---}
The normal state of the underdoped high temperature superconductors
(SCs) has long been argued to lie outside the paradigm of Fermi
liquid theory due to strong electron correlations. The most remarkable
manifestation of this appears in recent angle
resolved photoemission spectroscopy (ARPES) studies \cite{fermiarcs}
above the superconducting transition temperature ($T_c$) in underdoped (UD)
Bi$_2$Sr$_2$CaCu$_2$O$_{8+\delta}$. These experiments show that the 
spectral weight of low energy single particle excitations is not present 
on a closed Fermi surface (FS) in momentum space. Instead, it forms open 
``Fermi arcs'' \cite{fermiarcs} of length $\sim T/T^*$, where $T^*$ is the
pseudogap temperature, indicating a complete breakdown of the Fermi
surface between $T_c$ and $T^*$. 
One can envision two very different causes for such NFL behavior.

(i) The ARPES experiments could be probing the finite temperature
properties of some exotic non-SC ground state which competes with
d-wave superconductivity at low doping.  Since the Fermi arcs appear 
to extrapolate to nodal points enclosing zero volume \cite{fermiarcs} 
as $T \to 0$, this
``underlying normal state'' must be a non-Fermi liquid as it
violates Luttinger's theorem \cite{luttinger,oshikawa,z2luttinger}. 
Varma and Zhu \cite{varma} have
postulated a time-reversal symmetry breaking phase with circulating
currents and argued that current fluctuations in this phase can
lead to Fermi arcs. Other proposals for candidate underlying normal
states include the d-density wave \cite{ddw} and staggered flux phases 
\cite{sflux}.
With varying doping, these two states support hole pocket Fermi surfaces
with an area proportional to the doping,
although the spectral weight around this pocket is anisotropic and
has been argued to lead to an arc-like feature. Another
concern with these two states is that the ``arcs'' in such models do not 
appear to
extrapolate into nodal points as $T \to 0$. One more recent proposal
for arcs involves inducing an additional nematic liquid crystalline 
order in the SC state at low doping \cite{eakim} but so far it
has not addressed the normal state above $T_c$. 

\begin{figure}
\includegraphics[width=3.2in]{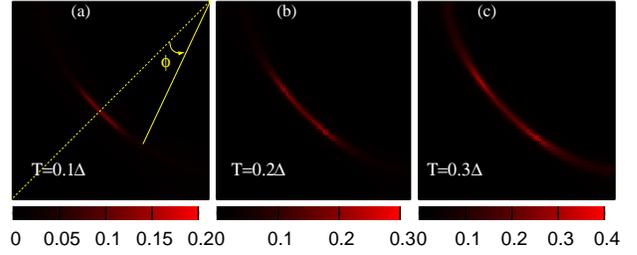}
\caption{(color online) Temperature evolution of Fermi arcs, regions
of significant $A(\bk,\omega=0)$ in the first quadrant of the
Brillouin zone.} \label{fig1} \vskip -4mm
\end{figure}

(ii) A second possibility is to examine whether order parameter fluctuations 
of the d-wave SC or fluctuations of the short range antiferromagnetic 
order above $T_c$ could lead to Fermi arc physics. This has been examined
earlier using various approaches \cite{otherarcs}, although these 
works have not focussed on the temperature scaling of the arc length.
A distinct way to study thermal order parameter fluctuations is to 
consider the effect of a dilute gas of classical vortices \cite{altman} 
above $T_c$. Alternatively, at low dopings and temperatures, quantum
fluctuations of the order parameter are also relevant; such quantum
fluctuations could be particularly important if one is close to a
quantum phase transition out of the superconducting state into a
correlated insulating state.

In this paper we will explore this last 
route. Namely, we study the possibility that the Fermi arcs in the 
underdoped normal state arise from proximity to a quantum phase transition 
between a d-wave SC and a correlated insulating state at low doping. 
Fig.~1 illustrates one of the key results of this paper --- the computed 
zero energy spectral function for the 
model which we study in this paper shows temperature dependent
Fermi arcs in the normal state as seen in ARPES.

There are several experimental motivations to explore this route which 
we follow. First, the observed gap structure in the normal state 
appears to extrapolate smoothly with lowering temperature to a SC gap 
with a pure d-wave form over a range of dopings. This suggests that any
``underlying normal state'' scenario has to be specifically chosen to
extrapolate into the excitation spectrum of the d-wave state over a 
whole range of dopings. This seems, to us, to be slightly unnatural
although we cannot rule it out. It appears more natural to
assume that the normal state has fluctuating d-wave correlations. Second, 
the arc length seen in the
experiments appears to scale linearly with temperature and independent
of most microscopic details. Such a simple $T$-dependence suggests 
proximity to a quantum phase transition near which the temperature 
naturally arises as the only energy scale. Finally, NFL behavior
has been observed in a variety of other 
materials \cite{sr3ru2o7,ybrh2si2} near field or pressure tuned
quantum phase transitions. Given this ubiquity of NFL behavior in the
vicinity of quantum phase transitions, it is worth examining this
scenario also for the underdoped cuprates.
Our proposal receives support from 
penetration depth measurements\cite{broun05} on
YBa$_2$Cu$_3$O$_{6+x}$ which appear to be probing the low
temperature manifestations of a 3D SC-insulator quantum phase transition 
\cite{herbut05,iyengar06}. Recent tunneling experiments \cite{davis}
in other
cuprates also show that the SC state at low hole density is destroyed 
by charge localization leading to such a SC-insulator transition.

{\it SC-insulator transition, critical charge fluctuations and the
electron spectral function. ---}
The underdoped cuprate SCs are doped correlated insulators, 
and the strongly correlated Hubbard model is thought to be a possible
minimal description of this system \cite{lee-rmp}. 
At large $U$, this leads to
an effective $tJ$ model with a kinetic energy for the doped holes
and a superexchange interaction between the spins. 
In order to calculate the electron spectral function in such a system, 
we appeal to slave-boson mean field theory
\cite{kotliarliu,bza,senthilz2,lee-rmp} which decomposes the electron 
operator into a product of a spin-1/2 neutral spinon and a
spinless charge-1 boson, $c^\dagger_\sigma (\br,\tau) \sim
f^\dagger_\sigma(\br,\tau) \Psi(\br,\tau)$. These two particles
are coupled by gauge fields enforcing the non-double occupancy constraint. 
At mean field level \cite{kotliarliu,lee-rmp}, the spinons and charges
are decoupled since gauge fluctuations are ignored --- 
the SC state then corresponds to d-wave pairing of spinons
and condensation of the bosons. Despite this drastic 
approximation, the SC correlations obtained in this theory are in 
remarkable qualitative agreement with a careful treatment using
Gutzwiller projected wavefunctions. This wavefunction approach
works directly with electronic
degrees of freedom, and has been shown to
{\it quantitatively} explain a variety of photoemission and optical
conductivity data \cite{paramekanti} in the SC state. At finite
temperature, gauge fluctuations around the mean field state are
more important; however their effect has been shown to be
mitigated in the presence of critical boson fluctuations \cite{ybkim}
(which we are interested in here), although the gauge fluctuation corrections
to the results presented here remains an important problem for future 
work. For now, we will therefore view slave boson mean field theory as 
a crude but qualitatively reasonable way to proceed, in order to see 
what aspects of the phenomenology of the underdoped cuprates can be
possibly captured.

With underdoping, the SC undergoes a quantum phase
transition into a correlated insulating state which, as suggested by
tunneling experiments \cite{davis}, has glassy bond-centered charge 
order which could arise from the Coulomb interactions between the bosons. 
However, in order to expose the physics of the Fermi arcs we choose
to study a simpler model of a SC-insulator transition, where the
boson field $\Psi(\br,\tau)$ is described by a Ginzburg Landau action
\be S_{\Psi}\!\!=\!\! \int d\tau d^3\br \left(\!
|\partial_\tau \Psi|^2 \!+\! c^2 |\nabla\Psi|^2\! +\! m^2 |\Psi|^2
\! +\! \frac{g}{2} |\Psi|^4 \right). \label{S} \ee Note the SC is
three-dimensional and anisotropic, $c^2 |\nabla\Psi|^2$ is the short hand 
notation for $c_{\parallel}^2 |\nabla_{\parallel}\Psi|^2+c_{z}^2
|\partial_z\Psi|^2$ , with $c_{z}\ll c_{\parallel}$. For $m^2 > 0$,
the system is in a uniform insulating phase with a charge gap. For $m^2 <
0$, the SC state with a nonzero $\la \Psi \ra$ is more stable. The
mean field SC-insulator critical point is at $m^2 = 0$. Model
(\ref{S}) is at its upper critical dimension and we can drop the
quartic interaction at $T=0$. The critical boson Green function is
then $\chi(\bQ,i\nu_n)=\la \Psi^*_{\bQ,i\nu_n} \Psi_{\bQ,i\nu_n} \ra
= 1/(\nu^2_n + \Omega^2_\bQ)$, where $\bQ\equiv(\bq,q_z)$ is the 3D
momentum with $\bq$ being its ab-plane component, $\Omega^2_\bQ =
c^2_{\parallel}\bq^2+c_z^2q^2_z$, and $\nu_n=2n\pi T$. Sitting at
$m^2=0$ and warming up from zero temperature, the quartic interaction 
induces \cite{subir}
a $T$-dependent mass $R$ leading to the charge
fluctuation propagator
\be
\chi^{-1}(\bQ,i\nu_n) = \nu^2_n + \Omega^2_\bQ + R 
\label{bosonG}
\ee 
Leading order
perturbation theory in $g$ at 1-loop (Hartree) level yields
$R=4g\sum_{\bQ}\Omega^{-1}_\bQ n(\Omega_\bQ)$, where $n(x) = 1/({\rm
e}^{x/T}-1)$ is the Bose function. At finite temperature, the
interaction also induces a damping term in the boson Green function.
This thermal relaxation rate is however higher order in $g$ in a
perturbative treatment, and we ignore it here. A description of 
the thermal relaxation rate over a range of energies from small to 
large values of $\nu/T$ is nontrivial \cite{subir}; we have so far not 
pursued this calculation.

We assume the spinons are paired in the d-wave channel throughout the 
transition, and described by a d-wave Bardeen-Cooper-Schrieffer
type Hamiltonian with a kinetic energy
$\xi_\bk = -2 t (\cos k_x + \cos
k_y) - 4 t' \cos k_x \cos k_y - \mu$, and a d-wave pairing gap
$\Delta_\bk = \Delta (\cos k_x - \cos k_y)/2$,
with the resulting quasiparticle dispersion $E_\bk =
\sqrt{\xi_\bk^2 + \Delta^2_\bk}$. 
Support for this approximation comes from thermal 
transport
measurements \cite{sutherland05}
which indicate that the thermal conductivity, which is dominated
by nodal d-wave quasiparticles in the SC state, {\it remains unchanged when we
enter the non-superconducting state below a critical doping}. Using this,
the electron Green function can be obtained by convolving its spinon and 
boson parts,
\be
G(\bk,ik_n)\!=\!\sum_{\bQ}
\frac{F_1\, u^2_{\bk-\bq}  + F_2 \, v^2_{\bk-\bq}}{2 N \omega_\bQ}.
\label{G}
\ee
where $N$ is the number of sites,
\bea
F_1&=&
\frac{1+n(\omega_\bQ)-f(E_{\bk-\bq})}{ik_n-E_{\bk-\bq}-\omega_\bQ}
+
\frac{n(\omega_\bQ)+f(E_{\bk-\bq})}{ik_n-E_{\bk-\bq}+\omega_\bQ},
\nonumber
\\
F_2&=& \frac{1+n(\omega_\bQ)-f(E_{\bk-\bq})
}{ik_n+E_{\bk-\bq}+\omega_\bQ} +
\frac{n(\omega_\bQ)+f(E_{\bk-\bq})}{ik_n+E_{\bk-\bq}-\omega_\bQ}.
\nonumber
\eea
Here $\omega_\bQ=\sqrt{\Omega^2_\bQ+R}$, $f(x) = 1/({\rm
e}^{x/T}+1)$ is the Fermi function, and
$u^2_\bk=(1+ \xi_\bk/E_\bk)/2$ and
$v^2_\bk=1-u^2_\bk$ are the usual BCS coherence factors.
Our work is quite close in spirit 
to the zero temperature calculations of Ref.~\cite{lannert}.

After analytic continuation, $ik_n\rightarrow \omega+i0^{+}$, the imaginary 
part of Eq. (\ref{G}) gives the electron spectral function
$A(\bk,\omega)$.  The result is easy to understand 
qualitatively.
An electron with energy $\omega$ and in-plane momentum $\bk$ is made by 
combining a spinon and a boson. However, the spinon Green function has
weight at energies $\pm E_{\bk-\bq}$ and the boson excitations also has 
weight at $\pm \Omega_\bQ$ (as seen from the boson Green function in
Eq.~\ref{bosonG}) which leads to four contributions to the electron
spectral function.
The thermal and coherence factors appearing in the spectral function 
appropriately weight the probability of such processes.


{\it Fermi arcs in $A(\bk,0)$. --- } 
We first argue that Eq.~(\ref{G}) leads to Fermi
arcs centered around the nodes of the d-wave SC dispersion if we
focus on the zero energy spectral function $A(\bk,0)$. In order to 
make the qualitative argument, let us work 
at very low temperatures such that $T \ll \Delta, c_\parallel \Lambda,
c_\perp \Lambda$. In this low temperature
regime, the  interaction induced mass $R \sim T^2$. Thus, the
characteristic boson fluctuations will have very small in-plane
momenta $|\bq| \sim (T/c_\parallel) \ll \Lambda$, small overall
momentum $|\bQ| \ll \Lambda$, and energies $\sim T$.
In order to obtain a zero energy electron with a certain momentum $\bk$, 
we have to use the energy of this thermal boson fluctuation 
to excite a spinon quasiparticle, namely spinon quasiparticles with
binding energy $\sim T$. The zero energy electron will have nearly
the same in-plane momentum as the spinon quasiparticle (since the 
boson carries very little momentum). Given these constraints, 
and the fact that spinon quasiparticles with energy $\sim T$ lie on
banana shaped
arcs of size $\sim T/\Delta$ around the d-wave nodes of the SC, it is 
clear that the zero energy electrons will also lie on ``Fermi 
arcs'' of size $T/\Delta$ around the nodes of the underlying d-wave
SC. Since $T^* \sim
\Delta$ in the cuprates, this is consistent with $(T/T^*)$
scaling of Fermi arcs reported in Ref. \cite{fermiarcs}.

\begin{figure}
\includegraphics[width=3.4in]{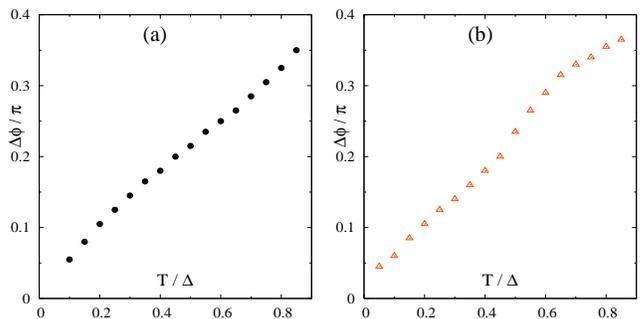}
\caption{(color online) Estimate for the arc length $\Delta\phi$
from the
zero energy spectral weight $A(\phi,0)$ for angles $\phi$ along
the FS (defined in Fig.~1).
(a) Arc length defined as angular distance between points of
maximum $A(\phi,0)$ along the Fermi surface. (b) Arc length
defined as the region with half the total spectral weight along
the FS. See text for details.} \label{fig2}
\vskip -4mm
\end{figure}

The above rough argument relied on working at very low temperature. 
We next proceed to a numerical evaluation of the electron spectral
function $A(\bk,0)$ in order to justify this argument and also
to study the evolution upto temperatures $T \sim \Delta$.
We have numerically evaluated $A(\bk,0)$ using Eq.(\ref{G}) for
temperatures of interest to the ARPES experiments ($T = 0.1\Delta -
0.9\Delta$). We work at doping $\delta=0.05$ and choose
$t'/t=-0.25$, $\Delta/t=0.15$ for the spinons. Also we set
$c_{\parallel}\Lambda = 3t \gg \Delta$, $c_z=c_{\parallel}/300$, and
use a Lorentzian broadening of $0.01t$ in Eq.(\ref{G}). The results for 
$g=0.1$ are plotted in Fig.~1 for $0<k_x,k_y<\pi$, qualitatively similar
results are found for other values of $g$. We see from Fig.1(a-c)
that $A(\bk,0)$ forms open arcs in momentum space with negligible
(although nonzero) weight at other momenta on the Fermi surface.

{\it Size of the Fermi arcs. --- }
In order to measure the size of the Fermi arcs, we
analyze the zero energy momentum distribution
curve along the spinon Fermi surface, $A(\phi,0)$, where $\phi$ is
the angle from the (0,0)-($\pi,\pi$) diagonal as shown in Fig.1(a).
Going from the nodal towards the antinodal direction along the
spinon Fermi surface, $A(\phi,0)$ first reaches its peak value at
$\phi_{max}$ and then drops to a minimum (background) value of
$A_{bg}$ at $\phi=\pi/4$. We use $\Delta\phi=2\phi_{max}$, i.e.
the angular span between two spectral peaks along the Fermi surface,
as a first measure of the arc size. Its approximately linear 
$T$-dependence is shown in Fig.~2(a).

An alternative measure for the arc size is given by
$\Delta\phi=2\phi_h$, where the region $0<\phi<\phi_h$ accounts for
half the total integrated spectral weight, $\int_0^{\phi_h} d\phi
[A(\phi,0)-A_{bg}]=\frac{1}{2}\int_0^{\pi/4} d\phi
[A(\phi,0)-A_{bg}]$. $\Delta\phi$ defined in this way is shown
in Fig.~2(b). We see that both definitions yield the
similar qualitative result that $\Delta\phi$ scales linearly with $T$.
This is consistent with our earlier arguments.
(The quantitative agreement between the two definitions
above is sensitive to the details of what percentage of the
spectral weight is chosen in the second definition.)

While the above two definitions of the arc length
are accessible theoretically,
they are harder to measure in experiments due, atleast, to momentum 
dependent matrix elements. We therefore turn to results for the
EDC at fixed momentum.

\begin{figure}
\includegraphics[width=3.4in]{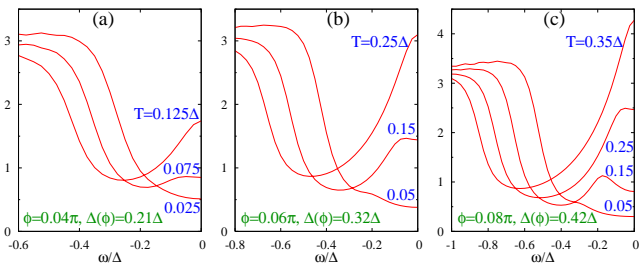}
\caption{(color online) Energy distribution curve $A(\phi,\omega)$
for various angles $\phi$ along the FS, and for increasing temperature
showing the filling in of the low energy spectral weight. For angles
closer to the node as in (a), the filling in of the spectral weight occurs 
at lower temperatures consistent with experiments. See text for discussion.} 
\label{fig3}
\vskip -4mm
\end{figure}

{\it Energy distribution curve (EDC) lineshape. ---}
Fig.~3 shows the EDC, namely $A(\phi,\omega)$, plotted
over a range of temperatures
and angles along the Fermi surface. At all angles, we find that there
is a filling-in of low energy spectral weight as the temperature
is raised. In our model, this spectral weight
growth is facilitated by the temperature dependent mass $R$ which
transfers weight from high to low energies.
The recovery of zero energy spectral weight
occurs at lower temperature for near-nodal points (smaller $\phi$)
than at angles closer to the antinodal direction. This growth of the 
low energy spectral weight is such that the dip-like EDC at low energy 
evolves into a
peak-like structure which at sufficiently high temperature becomes a
dominant feature. 

The above results are broadly consistent with the experimental
observations \cite{fermiarcs}. While the detailed EDC lineshape is
not quite like the experimental lineshape, the comparison
is expected to improve if we take into account the following
points which have been ignored in our analysis. (i)
The shift of the higher energy edge-like feature
to larger $\omega$ could be somewhat offset by
a temperature-dependent spinon pairing gap as the local singlet
correlations weaken with increasing $T$ --- this needs a more
careful self-consistent calculation of the spinon gap. (ii) The
low energy lineshape may improve if relaxational dynamics for the
bosons \cite{subir} is taken into account. (iii)
Our model Eq.(\ref{G}) also ceases to be a
good approximation at large energies where other inelastic
channels, including spin fluctuations, become important. (iv) The
nonzero energy resolution of the ARPES experiments ($\sim 15-30 meV$)
has not been taken into account here. Given these
limitations, we have not attempted a more detailed analysis
to extract the arc length from the EDC as in the experiments.

{\it Discussion. ---}
We have explicitly shown, within a phenomenological slave-boson approach, 
that critical charge fluctuations near a SC-insulator transition as 
described by Eq. (\ref{S}), together with spinon pairing in the d-wave 
channel, gives rise to Fermi arcs with size $\sim T/T^*$. We finally
present some arguments for how the
normal state of weakly doped cuprates probed in ARPES 
experiments \cite{fermiarcs} could lie in the quantum critical regime of
such a transition.
Recent tunneling experiments \cite{davis}
in the weakly doped cuprates indicate that 
the SC state at low doping undergoes a charge localization into an
insulating state. Such a charge localization transition, where the
doped holes become immobile, is consistent with a SC-insulator between 
a d-wave SC and an insulating state which retains d-wave paired spinons.
Some hints for the existence of d-wave nodal spinons at low
temperature comes from thermal conductivity measurements \cite{sutherland05}
in samples
with $T_c=0$, which show a residual $\kappa/T$ very close to that of 
the SC state at higher doping. This scenario for the SC-insulator
transition is sketched in Fig.~4. Warming up above $T_c$ in the low doping 
regime could thus land us in the quantum critical regime of a 
SC-insulator transition. Although we have not specifically studied this
transition from the SC to a glassy charge-localized insulator with 
d-wave spinons, we
believe our results strongly suggest that this mechanism of critical
charge fluctuations near a SC-insulator transition could lead to
$T$-linear Fermi arcs in the underdoped normal state above $T_c$.
It would be interesting to look for more direct experimental signatures 
of such quantum fluctuations in the underdoped normal state.

\begin{figure}
\includegraphics[width=2.2in]{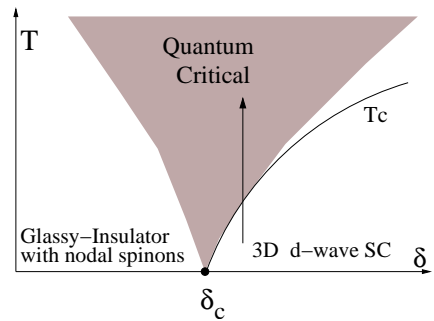}
\caption{(color online) Tunneling and thermal conductivity measurements
suggest that the cuprate SCs evolve, with doping, from a glassy charge 
insulator with nodal spinon excitations into a 3D d-wave SC.
In the weakly doped regime above $T_c$, we expect quantum charge
fluctuations and Fermi arcs similar to that in the simple model studied 
here.} 
\label{fig4}
\vskip -3mm
\end{figure}

We thank E. Altman, J. C. Davis, Y.-B. Kim, P. A. Lee, M. Norman and
a referee for comments and discussions. We are particularly indebted 
to A. Vishwanath for valuable suggestions on an earlier version of this 
paper. We acknowledge support from NSERC (AP,EZ) and the A. P. Sloan 
Foundation (AP).

\end{document}